\documentclass[12pt]{article}
\pdfoutput=1
\usepackage[a4paper]{geometry}
\usepackage{jcapmod}
\usepackage{amssymb}
\usepackage{bm}
\usepackage{latexsym}
\usepackage{amssymb}
\usepackage{amsmath}
\usepackage{amsfonts}
\usepackage{multirow}
\usepackage{array}
\usepackage{booktabs}
\usepackage{rotating}
\usepackage{graphicx}
\usepackage{color}

\def\clap#1{\hbox to 0pt{\hss#1\hss}}

\newcommand{\be}{\begin{equation}}
\newcommand{\ee}{\end{equation}}
\newcommand{\bea}{\begin{eqnarray}}
\newcommand{\eea}{\end{eqnarray}}

\newcommand{\EulerGamma}{\gamma_{\mathrm{E}}}

\DeclareMathOperator{\Or}{\mathcal{O}}

\newcommand{\half}{\frac{1}{2}}
\newcommand{\intg}{\int d^4x \sqrt{-g}}

\renewcommand{\vec}[1]{\bm{\mathrm{{#1}}}}

\newcommand{\para}[1]{\par\vspace{2mm}\noindent\emph{{#1}}.---}

\newsavebox{\tableA}
\newsavebox{\tableB}

\newsavebox{\boxplot}
\newsavebox{\boxplota}

\notoc

\definecolor{orange}{rgb}{1,0.5,0}

\begin{document}
\begin{flushright}
DAMTP-11-87 \\ MIFPA-11-48\\ NSF--KITP-11-218
\end{flushright}

	\title{The Importance of Slow-roll Corrections During Multi-field Inflation}

	\author[a,b]{Anastasios Avgoustidis,}
	\author[a,c]{\!Sera Cremonini,}
	\author[a]{\!Anne-Christine Davis,}
	\author[a]{Raquel H. Ribeiro,}
	\author[d]{Krzysztof Turzy\'{n}ski,}
	\author[e]{Scott Watson}

	\affiliation[a]{Department of Applied Mathematics and Theoretical Physics\\
	Centre for Mathematical Sciences, Wilberforce Road \\
	Cambridge CB3 0WA, United Kingdom}

      \affiliation[b]{School of Physics and Astronomy, University of Nottingham, University Park\\
      Nottingham NG7 2RD, United Kingdom}

	\affiliation[c]{George and Cynthia Mitchell Institute for Fundamental Physics
	 and Astronomy, \\ Texas A \& M University, \\College Station, TX 77843Ð4242, USA}

	\affiliation[d]{Institute of Theoretical Physics, Faculty of Physics, University of Warsaw, \\
	ul. Ho\.{z}a 69, 00-681 Warsaw, Poland}

	\affiliation[e]{Department of Physics, Syracuse University, \\
	Syracuse, NY 13244, USA}

	\emailAdd{A.Avgoustidis, S.Cremonini, A.C.Davis, R.Ribeiro@damtp.cam.ac.uk}
	\emailAdd{turzyn@fuw.edu.pl}
	\emailAdd{gswatson@syr.edu}

	\abstract{       We re-examine the importance of slow-roll corrections during the evolution of cosmological perturbations in models of multi-field inflation.  We find that in many instances the presence of light degrees of freedom leads to 
        situations in which next to leading order slow-roll corrections become significant. Examples where we expect such corrections to be crucial include models in which modes exit the Hubble radius while the inflationary trajectory undergoes an abrupt turn in field space, or during a phase transition.  
     We illustrate this with several examples -- hybrid inflation, double quadratic inflation and double quartic inflation. Utilizing both analytic estimates and full numerical results, we find that corrections can be as large as 20\%.  Our results have implications for many existing models in the literature, as these
corrections must be included to obtain accurate observational predictions --
particularly given the level of accuracy expected from CMB experiments such as
\textit{Planck}.    
	}


	\maketitle

\section{Introduction}
	\label{sec:introduction}

The inflationary paradigm \cite{Guth:1980zm,Linde:1981mu} remains a convincing causal mechanism for providing the needed initial conditions of the early universe -- despite scrutiny from a wealth of precision cosmological observations \cite{Komatsu:2010fb, Larson:2010gs}  (see \cite{Lidsey:1995np, Lyth:2009zz} for reviews).  Inflation accomplishes this by providing a period of quasi-de Sitter expansion that leads to a classical spectrum of nearly scale invariant density fluctuations \cite{Mukhanov:1981xt}. The evolving density contrast then leads to the large scale structure and Cosmic Microwave Background (CMB) anisotropies observed today.
In the simplest models of inflation the acceleration is assumed to be the result of a single scalar degree of freedom -- the inflaton.  In such models it was shown long ago that the perturbation in the spatial curvature is a conserved quantity on super-Hubble scales \cite{Bardeen:1983qw}, which makes single field models of inflation rather easy to analyze.

Embedding the concept of inflation in a more fundamental theory generically leads to the existence of additional light degrees of freedom.  
These fields typically influence the dynamics and may even help drive inflation, leading to multi-field models 
(early examples are \cite{Kofman:1986wm,Silk:1986vc}).
In such situations the curvature perturbation does not necessarily remain constant on super-Hubble
scales, resulting in new theoretical challenges as well as richer possibilities for 
observations \cite{Nakamura:1996da, Gordon:2000hv, GrootNibbelink:2001qt,
	Bernardeau:2002jy, DiMarco:2005nq, Byrnes:2006fr, Lalak:2007vi,
	Langlois:2008mn,Cremonini:2010sv}.
One example arises when the inflationary trajectory deviates from a geodesic in the field space \cite{Achucarro:2010jv, Achucarro:2010da}.  
As a result, the curvature perturbations are not constant after crossing the Hubble radius, and the na\"ive single-field picture fails.
Thus, the usual \emph{single-field} relations between observable quantities
(such as the amplitude of the perturbations and the spectral index inferred
from the two-point correlation function) and the parameters of the inflationary potential are lost.
Another example is when a time dependent coupling between fields approaches the regime of strong coupling,
which in turn can lead to interesting observational signatures \cite{Achucarro:2010jv, Cremonini:2010ua, Achucarro:2010da, Chen:2011zf, Baumann:2011su, Shiu:2011qw}.
	
Among the theoretical obstacles facing multi-field inflation models is finding simple (and universally applicable) tools 
for calculating the power spectrum of curvature perturbations, particularly in the situation where couplings between 
fields are significant.  A common and useful tool for analyzing multi-field models is the so-called $\delta N$ 
formalism \cite{Starobinsky:1986fxa,Sasaki:1995aw,Sasaki:1998ug},  
which allows to calculate the power spectrum of the curvature perturbations
to leading order in the gradient expansion and to all orders in the slow-roll parameters.
Indeed, this approach has led to interesting
and quite restrictive bounds on multi-field inflation models from primordial non-gaussianity \cite{Suyama:2007bg, Suyama:2010uj, Sugiyama:2011jt, Smith:2011if}.
However, using this method in practice
requires (typically reasonable) assumptions about the amplitude of the fluctuations at Hubble radius crossing.
In particular, one assumes that the amplitude is equal to that of perturbations of a light scalar field in pure de Sitter space, 
where leading order in slow-roll is more than adequate to capture the dynamics -- an intuitive and very accurate approximation in many cases.
However, 
in special cases of hybrid inflation this approximation may fail,
specifically when mildly tachyonic isocurvature modes source the curvature perturbations
\cite{Clesse:2010iz,Abolhasani:2010kn,Kodama:2011vs}.  
These special situations provide examples
where the leading slow-roll approximation can fail 
 to provide accurate initial conditions for
the $\delta N$ formalism.

With these motivations in mind we revisit multi-field models of
inflation.  We find the next to leading order corrections to the
power spectra of curvature and isocurvature perturbations near
Hubble radius crossing and on super-Hubble scales utilizing
the transfer matrix formalism.
We give simple examples of models where these corrections
are essential for obtaining accurate estimates of the power spectra.
This generalizes the work of 
Refs. \cite{Stewart:1993bc, Gong:2001he, Choe:2004zg, Chen:2006nt, Burrage:2011hd} to the multi-field case, and extends the results of Refs. \cite{Byrnes:2006fr,Peterson:2010np} to include higher order corrections and an explicit
expression for the transfer matrix of perturbations. 	

\para{Outline}%
The paper is organized as follows.
In \S \ref{sec:model} we review canonically normalised two-field inflation,
discuss the usual slow-roll approximations and motivate considering higher order slow-roll effects.
In \S \ref{sec:2pf} we	study the dynamics of the perturbations on a Friedmann--Robertson--Walker background,
around Hubble radius crossing.
We derive the expressions for the power spectra
	of curvature and isocurvature perturbations up to
	next-order in the relevant slow-roll parameters.
	These analytical results are valid up to a few e-folds after horizon crossing
	(when the modes have become classical), and capture the relevant time
	and scale dependence of the power spectra.
	The evolution of the perturbations on super-Hubble scales
	is the subject of \S\ref{sec:2pfb},
	where we calculate the transfer matrix for the perturbations
	up to next-next-order in slow-roll parameters.
	In \S\ref{sec:numerics} we present two simple but instructive examples
	of inflationary trajectories in hybrid inflation
	and double quadratic inflation models, for which the importance of the
	next-order corrections is manifest.
	We conclude in \S \ref{sec:conclusions}.
The appendix collects additional formulae necessary to compute the power spectra.

\section{Perturbations in multi-field inflation}
\label{sec:model}
We will restrict our attention to the case of two scalar fields minimally coupled to gravity, although our analysis is easily generalizable to more fields. At the level of renormalizable interactions the action is then
	\be
	\label{eq:action}
	  S=\intg \left[ \half \, m_p^2 R - \half \left( \partial \varphi \right)^2 -
	  \half \left( \partial \chi \right)^2 - V(\varphi, \chi) \right] \, ,
	\ee
	where  $m_p=1/\sqrt{8 \pi G} \simeq 2.43 \times 10^{18}$ GeV is the reduced Planck mass.
	The background equations of motion for the fields $\phi = \{\varphi,\chi\}$ are given by
	\be
	\label{eq:background_eom}
	\ddot{\phi}+3H\dot{\phi}+V_{,\phi}=0  \, ,
	\ee	
   where
   the dot denotes differentiation with respect to cosmic time
   and
   $V_{,\phi}\equiv\partial V/\partial \phi$.
   The cosmological evolution is described by the Friedmann equations
   	\begin{subequations}
	\begin{align}
	\label{eq:friedmann}
	3H^2 m_p^2 & =\dot{\varphi}^2+\dot{\chi}^2+V(\varphi, \chi)\, , \\
	 2\dot{H} m_p^2&=-\big[\dot{\varphi}^2+\dot{\chi}^2 \big] \, ,
	\end{align}
	\end{subequations}
	where $H\equiv \dot{a}/a$ is the Hubble parameter and in what follows we will work in units where $m_p=1$ for simplicity.
	We now expand the fields in fluctuations around their homogeneous background values
	\begin{subequations}
	\begin{align}
	\varphi(t,\vec{x}) &=\varphi_0(t)+\delta\varphi(t,\vec{x})\, , \\
	\chi(t,\vec{x}) &=\chi_0(t)+
	\delta\chi(t,\vec{x}) \, .
	\label{eq:perturb}
	\end{align}
	\end{subequations}
A particularly convenient basis for interpreting the behavior of the fluctuations \cite{Gordon:2000hv} is found by performing the instantaneous rotation
         \begin{subequations}
         \begin{align}
	\delta \sigma & \equiv \cos\theta \,\delta \varphi +\sin \theta \,\delta \chi \, ,\\
	\delta s & \equiv -\sin\theta \, \delta \varphi +\cos \theta \, \delta \chi \ .
	\end{align}
	\end{subequations}
	The rotation angle $\theta$ is given by
         $\cos\theta = \dot\varphi/\dot\sigma$ and $\sin\theta=\dot\chi/\dot\sigma$,
         where
	$\dot{\sigma} \equiv \sqrt{\dot{\varphi}^2+\dot{\chi}^2}$.
	The background equations of motion \eqref{eq:background_eom} then become
	\begin{subequations}
	\begin{align}
	& \ddot{\sigma}+3H\dot{\sigma}+V_{,\sigma}=0 \, ,\\
	&  \dot{\theta}=-\dfrac{V_{,s}}{\dot{\sigma}} \ ,
	\end{align}
	\end{subequations}
     where $V_{,\sigma}=\cos\theta\, V_{,\varphi} + \sin\theta\, V_{,\chi}$ and
	$V_{,s}=-\sin\theta\, V_{,\varphi} + \cos\theta\, V_{,\chi}$.\footnote{More generally,
	with $I,J \in\{\sigma,s\}$ and $i,j\in\{\varphi,\chi\}$, we have
	$V_{,IJ\ldots K}\equiv E_I^i E_J^j\ldots E_K^k V_{,ij\ldots k}$,
	where $E_\sigma^\varphi=E_s^\chi=\cos\theta$, $E_\sigma^\chi=
	-E_s^\varphi=\sin\theta$.}
From the background equations we see that in this basis $\delta \sigma$ corresponds to the scalar field fluctuations along the trajectory of the inflaton (adiabatic perturbations), whereas $\delta s$ is the fluctuation orthogonal to the trajectory (isocurvature perturbations).

Whereas the entropy perturbation $\delta s$ is automatically gauge invariant, we need to introduce a
gauge invariant definition for the adiabatic perturbations.  This may be accomplished by introducing the Mukhanov-Sasaki \cite{Mukhanov:1985rz, Sasaki:1986hm}
variable $Q_\sigma=\delta\sigma-\frac{\dot\sigma}{H}\Phi$, 
which corresponds to the instantaneous curvature perturbation on surfaces of constant $\sigma$, where $\Phi$ is the Bardeen 
potential describing scalar perturbations of the metric\footnote{On super-Hubble scales, $Q_\sigma$ is related to the 
comoving curvature perturbation via ${\mathcal{R}}=(H/\dot\sigma)Q_\sigma$. Similarly, we can define a comoving 
isocurvature perturbation through $\mathcal{S}=(H/\dot\sigma)\delta s$.}.

Given the gauge invariant perturbations $Q_\sigma$ and $\delta s$, it is convenient to introduce the rescaled variables 
$u_\sigma\equiv a\,Q_\sigma$ and $u_s\equiv a\,\delta s$, in terms of which 
the equations of motion for modes with comoving momentum $k$ take the form \cite{Lalak:2007vi}
	\begin{subequations}
	\begin{equation}
		u''_{\sigma}+2 a \dfrac{V_{,s}}{\dot{\sigma}} u'_{s} +
		\bigg[k^2 - \dfrac{a''}{a}+a^2 C_{\sigma\sigma}   \bigg]u_{\sigma}
		+ \bigg[-2a' \dfrac{V_{,s}}{\dot{\sigma}}+a^2 C_{\sigma s}  \bigg] u_s
		= 0
		\label{eq:eom_usigma}
	\end{equation}
	and
	\begin{equation}
	u''_{s}-2 a \dfrac{V_{,s}}{\dot{\sigma}} u'_{\sigma} +
		\bigg[k^2 - \dfrac{a''}{a}+a^2 C_{s\sigma}   \bigg]u_{s}
		+ \bigg[2a' \dfrac{V_{,s}}{\dot{\sigma}}+a^2 C_{ss}  \bigg] u_{\sigma}
		= 0 \ \ ,
		\label{eq:eom_us}
	\end{equation}
	\end{subequations}
	where primed quantities are differentiated
	with respect to conformal time $\tau \equiv\int_{}^{t} \mathrm{d}\tilde t/a( \tilde t)$
	and the coefficients $C_{IJ}$ are listed in Eqs.\eqref{eq:coo}.
We note that, since we are restricting our attention to classical scalar fields with renormalizable interactions, 
both the scalar fluctuations $u_{\sigma}$ and $u_s$ propagate at the speed of sound $c_s^2=1$.

\subsection{Slow-roll approximation: the need for next-order contributions}	
\label{sec:sr_diag}
	
	We can simplify the equations of motion for the perturbations
	 \eqref{eq:eom_usigma} and \eqref{eq:eom_us} 	
	by applying slow-roll conditions.
	In many realistic models of inflation,
	the approximate constancy of the Hubble parameter requires
	$\varepsilon\equiv-\dot H/H^2 \ll 1$ , while the
	approximate scale invariance of the power spectrum of perturbations
	at Hubble radius crossing (proportional to $H^2/\varepsilon$) requires the smallness of
	\be
	\frac{\dot\varepsilon}{\varepsilon H} = 2 (2\varepsilon-\eta_{\sigma\sigma})\, ,
	\ee
        where $\eta_{IJ}=V_{,IJ}/3H^2$. This yields $\varepsilon,|\eta_{\sigma\sigma}|\ll1$.
         These requirements, however, do not place severe constraints on the remaining
         slow-roll parameters, since the time evolution of $\eta_{IJ}$ is given by \cite{Byrnes:2006fr}:
         \begin{subequations}
         \begin{align}
         \label{eq:sr_der1}
         \frac{1}{H}\dot\eta_{\sigma\sigma} & = 2(\varepsilon\eta_{\sigma\sigma}-\eta_{\sigma s}^2)-\xi_{\sigma\sigma\sigma}^2 \;  , \\
         \frac{1}{H}\dot\eta_{\sigma s} & = \eta_{\sigma s}(2\varepsilon+\eta_{\sigma\sigma}-\eta_{ss}) -\xi^2_{\sigma\sigma s},  \;  \\
         \label{eq:sr_der3}
         \frac{1}{H}\dot\eta_{ss} & = 2(\varepsilon\eta_{ss}+\eta_{\sigma s}^2) - \xi^2_{\sigma ss} \; ,
         \end{align}
         \end{subequations}
	where $\xi^2_{IJK}\equiv V_{,\sigma} V_{,IJK}/V^2$.
The quantities	$3H^2\eta_{IJ}$ contribute to the mass matrix of the perturbations.
In particular, $\eta_{\sigma s}$ parametrizes the coupling between curvature and isocurvature perturbations
and it is therefore related to the bending of the inflationary trajectory in field space via
$\dot{\theta}/H=-\eta_{\sigma s}/(1-\epsilon/3)$.
	
As also observed in \cite{Shiu:2011qw}, there are three patterns for the slow-roll parameters $\varepsilon$ and $\eta_{IJ}$ consistent with the
	requirements mentioned above. One possibility is that they all have magnitudes much smaller than one. This
	is the pattern most often considered in the discussion of two-field inflationary models.
	Another possibility is that $\varepsilon,\,|\eta_{\sigma\sigma}|\ll1$, $\eta_{\sigma s}\simeq 0$
	and $\eta_{ss}$ is positive but otherwise arbitrarily large.  This is the decoupling regime, which is effectively
	equivalent to a single-field inflationary model.  A pattern interpolating between these
	two possibilities is also conceivable with
	$\left( \varepsilon,\,|\eta_{\sigma\sigma}|,\,|\eta_{\sigma s}| \right) \ll|\eta_{ss}|<1$.

	If all slow-roll parameters are much smaller than one, the
	evolution of the modes is simplified: the perturbations freeze-in near
	Hubble radius crossing, and the effect of a small and almost constant coupling
	between curvature and isocurvature modes can be easily included \cite{DiMarco:2005nq}.

	A quantitative description becomes more intricate if there is a significant
	super-Hubble evolution of the isocurvature modes, as in the recently considered
	models of hybrid inflation \cite{Clesse:2010iz,Abolhasani:2010kn,Kodama:2011vs},
    allowing for several tens of e-folds of inflation \emph{after} a
	mild tachyonic instability develops, associated with a moderate (and negative) $\eta_{ss}$.
	 One can, however, get some insight into the inflationary dynamics
	by solving Eqs.\ (\ref{eq:sr_der1})--(\ref{eq:sr_der3}) in the formal limit of $\varepsilon\to0$
	and with negligible $\xi^2_{\sigma IJ}$. With initial conditions $\eta_{\sigma\sigma}\approx0$,
	$\eta_{\sigma s}\approx0$ and $\eta_{ss}=\eta_{ss}^{(0)}<0$ at early times, the solution is:
	\begin{equation}
	\label{eq:sr_dersol}
	\eta_{\sigma\sigma,ss}=\frac{\eta_{ss}^{(0)}}{2}\bigg[1\mp\mathrm{tanh}\big[\eta_{ss}^{(0)}(N-N_0)\big]\bigg]\,,\qquad \eta_{\sigma s}=\pm\frac{\eta_{ss}^{(0)}}{2\mathrm{cosh}\big[\eta_{ss}^{(0)}(N-N_0)\big]}\, ,
	\end{equation}
	where $N$ is the number of e-folds and $N_0$ is a constant of integration. 
	Hence, although the hierarchy 
		$\left( \varepsilon,\,|\eta_{\sigma\sigma}|,\,|\eta_{\sigma s}| \right) \ll|\eta_{ss}|<1$
          is not stable if $\eta_{ss}<0$, it can persist for a large number of e-folds and 
          the transition to a stable one with $\eta_{\sigma\sigma}=\eta_{ss}^{(0)}$
          takes approximately $1/|\eta_{ss}^{(0)}|$ e-folds.
          It is therefore interesting to study the inflationary dynamics 
          with moderate and negative $\eta_{ss}$.

Another instance in which the dynamics can be quite involved
is the case when the slow-roll parameters are small only initially,
with some becoming of order one at a later
time, without violating the slow-roll condition $\varepsilon<1$.
This is the case, for example, in the	
model of double quadratic inflation~\cite{Langlois:1999dw}
or double quartic inflation.
With $\eta_{\sigma\sigma}=\alpha\varepsilon$, as it is the case for a single monomial dominating the inflationary potential, we can initially neglect $\eta_{\sigma s}$, $\eta_{ss}$, $\xi^2_{\sigma\sigma s}$ and $\xi^2_{\sigma ss}$
in Eqs. (\ref{eq:sr_der1})--(\ref{eq:sr_der3}), obtaining
\begin{equation}
	\label{eq:sr_dersol2}
\varepsilon=\frac{\varepsilon^{(0)}}{1-2\varepsilon^{(0)}(2-\alpha)(N-N_0)}\, , \qquad\eta_{\sigma s} = \eta_{\sigma s}^{(0)} \left(\frac{\varepsilon}{\varepsilon^{(0)}}\right)^{\frac{2+\alpha}{2(2-\alpha)}} \, ,
\end{equation}
which means that a sudden growth in $\varepsilon$ and $\eta_{\sigma\sigma}$ drives a 
subsequent growth
of $\eta_{\sigma s}$ and $\eta_{ss}$ until these slow-roll parameters backreact
on $\varepsilon$ and $\eta_{\sigma\sigma}$ and the approximation used here breaks down.

We emphasize that in all of these models, in order to describe accurately
the evolution of the modes outside the horizon,
one	must keep higher order terms in the slow-roll parameters.
This will be the focus of 
\S\ref{sec:numerics}.
Motivated by the discussion above, our strategy for probing the dynamics of the curvature and isocurvature modes will be the following.
First, in \S\ref{sec:2pf} we will track the evolution of the perturbations up to a few e-folds around Hubble radius crossing.
At this stage all slow-roll parameters are restricted to have magnitudes smaller than one, with the notable exception of $\eta_{ss}$.
As a result, the solutions of the equations of motion (\ref{eq:eom_usigma})--(\ref{eq:eom_us}) will have to be expanded
to next-order in $\eta_{ss}$, but only to leading-order in the remaining slow-roll parameters.
In \S\ref{sec:2pfb} we will then follow the evolution of the modes outside the Hubble radius.
Since in this region the term proportional to $k^2$ in Eqs.(\ref{eq:eom_usigma})--(\ref{eq:eom_us}) becomes negligible,
the solutions simplify significantly and enter the non-oscillatory regime.
However, as we will see, to obtain a faithful estimate of the amplitude of the perturbations
one generally needs to expand the equations of motion to next-order in \emph{all} slow-roll parameters.

\section{Dynamics near Hubble radius crossing}
\label{sec:2pf}

We start by noting that the expansions to next-order in slow-roll of the $C_{IJ}$ coefficients in
Eqs.\eqref{eq:coo} and of the background scale factor in Eqs.(\ref{eq:eom_usigma})--(\ref{eq:eom_us})
do not contain any terms proportional to $\eta_{ss}^2$. As a result, these
quantities can be expanded to leading-order in slow-roll.
The first order derivative terms in Eqs.\eqref{eq:eom_usigma} and \eqref{eq:eom_us}
can be removed by changing variables to
     \be
	\left( \begin{array}{c} w_1 \\ w_2 \end{array} \right) = \left( \begin{array}{c}
	           \cos \vartheta\, u_\sigma + \sin \vartheta\, u_s \\
	           -\sin \vartheta\, u_\sigma + \cos \vartheta\, u_s  \end{array} \right) \, ,
	           \label{wdef}
	\ee
where $\dot\vartheta=-V_{,s}/\dot\sigma$. As we justify in the Appendix,
the angle $\vartheta$ can be chosen so that the equations of motion for $w_1$ and $w_2$
decouple when a given reference scale, $k_\star$, crosses the Hubble
radius\footnote{Here and throughout the paper quantities with the subscript $\star$
are to be evaluated when the reference mode crosses the Hubble radius, \emph{i.e.} when $k_{\star}=a_{\star}H_{\star}$.}.
The equations of motion around the time of horizon crossing for $k_{\star}$ then simplify to
     	\be
	\left( \begin{array}{c}
					w''_{1}  \\
					w''_{2}
		   \end{array} \right) + \bigg[k^2 \mathbf{1} -\dfrac{1}{\tau^2}
		   										\left( \begin{array}{cc}
														2+\lambda_{1\star} & 0 \\
														0& 2+\lambda_{2\star}
									    	 			  \end{array} \right)
		     \bigg]
		  \left( \begin{array}{c}
					w_{1}  \\
					w_{2}
		   \end{array} \right)  = 0 \ \ ,
		\label{eq:eomw}
	\ee
where $\lambda_{1\star}$ and $\lambda_{2\star}$ are functions of the slow-roll parameters given in the Appendix.

\subsection{Slow-roll expansion of the mode functions}

The equations of motion \eqref{wdef} for the variables $w_A$
encapsulate the dynamics of the curvature and isocurvature perturbations around Hubble radius crossing.
The positive frequency solutions to the equations of motion \eqref{eq:eomw} are 
 represented, up to an irrelevant phase, by
\be
w_A \varpropto \dfrac{\sqrt{\pi}}{2} \sqrt{-\tau}\, H_{\mu_A}^{(1)} \,
	e_A(\vec{k})\,,
	\label{eq:wavefunction}
\ee
where $e_A(\vec{k})$ are Gaussian random 
variables\footnote{Since our goal is to calculate two-point 
correlation functions at tree level, 
the use of classical random fields here is equivalent to the proper 
procedure of expressing the Fourier modes in terms of creation and
 annihilation operators. Such a representation is legitimate on super-Hubble scales, 
 so this choice will allow us to make an easy comparison with the results of
  \S \ref{sec:2pfb}.} such that
$\langle e_A (\vec{k}) e_B^{*}(\vec{k}') \rangle =
	 \delta_{AB}\, \delta^{(3)}(\vec{k}-\vec{k}')$.
Here $A,B\in\{1,2\}$ and $H_{\mu_A}^{(1)} $ are Hankel functions of the first kind and of order
\be
\label{eq:mu}
\mu_A=\sqrt{\frac{9}{4}+3\lambda_A}\simeq \dfrac{3}{2}+\lambda_A-\dfrac{1}{3}\lambda_A^2\,.
\ee
We emphasize that the presence of $\lambda_A^2$ in the expansion (\ref{eq:mu}) introduces
terms proportional to $\eta_{ss}^2$ in the expansion of the Hankel function
in the slow-roll parameters.
We also note that deep inside the horizon, when $k\gg aH$, the modes are rapidly
oscillating and the solution \eqref{eq:wavefunction} reduces to the standard Bunch-Davies
vacuum \cite{Bunch:1978yq}. The two-point correlations of the $w_A$
variables satisfy
\be
\langle w_A^{\dagger} w_B \rangle = \dfrac{\pi}{4} (-\tau)
\big|H_{\mu_A}^{(1)} (-k\tau) \big|^2 \, \delta_{AB}\ \; ,
\label{eq:2pfw}
\ee
in terms of which we can write the correlation functions for
the $u_{\sigma}$ and $u_s$ perturbations \cite{Stewart:1993bc}.

Our goal is to obtain approximate expressions for the correlation functions in Eq.(3.5) around Hubble radius crossing. This exercise is particularly simple in single-field inflation models where the comoving curvature perturbation is conserved on super-horizon scales. There, the limit $|k\tau|\to 0$ gives both the correlation function value a few e-folds after horizon crossing and its asymptotic limit (with only the growing mode left) .
Indeed, if we take the decaying mode into account 3 to 5 e-folds after horizon crossing, the asymptotic limit of the Hankel function (with order approximately 3/2) will be corrected by $\mathcal{O}(10^{-6}-10^{-4})$ terms. In contrast, if inflation is driven by multiple fields
some of the slow-roll parameters might be non-negligible, which allows a significant
deviation of the order of the Hankel function from 3/2, which induces a much stronger time dependence than the presence of a decaying mode. In order to estimate this effect, we expand
the Hankel function up to second order
in the slow-roll parameters, keeping only the constant terms and the dominant contributions
when $|k\tau|\to 0$.
We find that 
\be
\Big| H_{\mu_A \star}^{(1)} (-k\tau) \Big|^2 \simeq -\dfrac{2}{\pi k^3 \tau^3}
			\bigg\{
			1-2\lambda_{A\star} \big[-2+\EulerGamma +\ln(-2k\tau) \big] +
			\lambda_{A\star}^2 f(-k \tau)
			\bigg\} \,,
			\label{eq:hf}
\ee
where $\EulerGamma\simeq 0.577$ is the Euler-Mascheroni constant and
the time dependent function $f(-k\tau)$ satisfies
\be
\begin{array}{ll}
6f (-k\tau) &= 16+3\pi^2+4(\EulerGamma+\ln2)(-11+3(\EulerGamma+\ln2))+ \\
&+(4+6(\EulerGamma-\ln2))\ln(-k\tau)+12\ln^2(-k\tau)\,.
\end{array}	
\ee
The discarded quadratic contributions in slow-roll parameters above are of the same order as the discarded decaying mode.
Therefore, this asymptotic expansion is valid up to $1$ part in $10^4$
and, as we will see in \S\ref{sec:2pfb}, it encodes the relevant dynamics needed for an accurate description
of the evolution of the modes outside the horizon.

Although the asymptotic expansion of the Hankel function in Eq.(\ref{eq:hf}) is often
a fairly accurate approximation of the predictions obtained by solving
the full equations of motion (\ref{eq:eom_usigma})--(\ref{eq:eom_us})
for the curvature and the isocurvature perturbations, it has some limitations.
 One of these comes from the expansion of the scale factor
\be
\label{eq:sca3}
a(\tau) \simeq -\dfrac{1}{H_{\star}\tau} \bigg[1+\varepsilon_{\star}
				+\varepsilon_{\star} \ln(-k_{\star}\tau) \bigg] \, ,
\ee
which is only valid for $|\ln(-k_{\star}\tau)|\lesssim 1/\varepsilon_{\star}$.
Secondly, the diagonal form of the mass matrix in Eq.(\ref{eq:eomw}) is strictly
true at Hubble radius crossing.  Since this matrix is rotated by an angle
$\sim \eta_{\sigma s}\ln(-k_{\star}\tau)$, our result is only limited to times for which
this angle can be considered small, unless $\lambda_{1\star}\simeq \lambda_{2\star}$.
Finally, as the slow-roll parameters evolve slowly in time, we can only treat them as
constants when $\ln(-k_{\star}\tau)$ is much smaller than any of the right hand sides
of Eqs.(\ref{eq:sr_der1})--(\ref{eq:sr_der3}).

\subsection{Power spectra}
The two-point correlation function of a given perturbation $Q$
is related to the (dimensionless) power spectrum via
\be
\langle Q_A(\vec{k})\, Q_B(\vec{k}') \rangle =(2\pi)^3 \delta^{(3)} (\vec{k}+\vec{k}')
\dfrac{2\pi^2}{k^3} \mathcal{P}_{AB}(|\vec{k}|)\ \ ,
\ee
where $Q_{A,B}={\mathcal{R}}, \mathcal{S}$ (repeated indices are customarily
shortened to a single one).
Evaluating the power spectra around the time of
horizon crossing, we find\footnote{This clarifies previous results found in Refs.\ \cite{Byrnes:2006fr, Lalak:2007vi}.}
		\begin{subequations}
	\begin{align}
		\mathcal{P}_{{\mathcal{R}}}  &
		= \dfrac{H_{\star}^2}{8\pi^2 \varepsilon_{\star}}
							\bigg\{1-2\varepsilon_{\star}
								+2\big[
									2-\EulerGamma -\ln2+\ln (k_{\star}/k)
									\big]
								 \big[
									3\varepsilon_{\star}-\eta_{\sigma\sigma\star}
									\big]
							\bigg\} \, , \label{eq:curv_spectrum}\\
		\mathcal{P}_{{\mathcal{R}} \mathcal{S}} & = \dfrac{H_{\star}^2}{4\pi^2 \varepsilon_{\star}}
				\bigg\{-2+\EulerGamma +\ln2+\ln (-k\tau) \bigg\}\, \eta_{\sigma s\star}\, , \\
		\mathcal{P}_{{\mathcal{S}}}  &
		 = \dfrac{H_{\star}^2}{8\pi^2 \varepsilon_{\star}}
				\bigg\{1-2\varepsilon_{\star} +2\ln(-k_{\star}\tau)
												\big[
												2\varepsilon_{\star} -\eta_{\sigma\sigma\star}+\eta_{ss\star}
												\big] \nonumber \\
					& \mbox{} 	\hspace*{1.8cm}
					+2(\varepsilon_{\star}-\eta_{ss\star}) (2-\EulerGamma -\ln2+\ln(k_{\star}/k)) +
					  \eta_{ss\star}^2 f(-k\tau)
				\bigg\} \  \label{eq:isocurv_spectrum},
			\end{align}
		\end{subequations}
		where $\mathcal{P}_{{\mathcal{R}}}$ denotes the power spectrum of the comoving
		curvature perturbation, $\mathcal{P}_{\mathcal{S}}$ that of the
		isocurvature perturbation, and $\mathcal{P}_{{\mathcal{R}} \mathcal{S}}$
		is the cross-correlation between the two perturbations.

Note that there are two types of logarithmic contributions in the power spectra:
those which depend explicitly on time, of the form $\ln(-k\tau)$, and those
which are scale dependent, $\sim \ln(k_\star/k)$.\footnote{The appearance of both families of logarithmic 
														contributions was mentioned in Ref. \cite{Burrage:2011hd}, 
														whereas Refs. 
														\cite{Gong:2001he, Zaldarriaga:2003my, Seery:2007wf}
														obtained time dependent logarithms.}
We can eliminate the latter type of logarithmic terms, by choosing an appropriate reference scale,
in this case  $k_{\star}=k$.
One still has to deal, however, with the time-dependent logs,
which \emph{cannot} be removed.
Although it follows from Eq.(\ref{eq:sca3}) that \emph{at} Hubble radius crossing
$-k\tau=1+\mathcal{O}(\varepsilon_\star)$
and the time-dependent logarithms vanish (up to corrections of higher order than
we are considering here),
we will see in \S\ref{sec:num_hyb_intro} that their inclusion beyond leading-order
may be essential for accurately tracing the evolution of the power spectra
for a few e-folds \textit{after} Hubble radius crossing.

Finally,
the absence of time-dependent terms in Eq.\eqref{eq:curv_spectrum}
is just a reflection of the fact that, without
isocurvature modes, the curvature perturbations are frozen after Hubble radius crossing.
We emphasize that the isocurvature modes will source the curvature perturbations,
but this is a next-order effect in the slow-roll parameters
not involving $\eta_{ss}$ and can therefore be neglected around Hubble radius
crossing, as we have argued previously\footnote{We expect $\mathcal{P}_{{\mathcal{R}}}$ to evolve as the
inflationary trajectory bends in field space. In particular, since $\dot\theta \simeq - H \eta_{\sigma s}$, its
logarithmic time
dependence will be proportional to $\eta_{\sigma s}^2$.}.
However, when describing the entire inflationary period this sourcing cannot be neglected,
as we will see explicitly in the next section.

\section{Dynamics after Hubble radius crossing}
\label{sec:2pfb}

The results derived in \S\ref{sec:2pf} are valid up to a few e-folds
after the relevant scales have exited the Hubble radius and the comoving
curvature perturbation $\mathcal{R}$ has become classical. To follow the subsequent
evolution of the modes, however, one needs to resort to other
techniques such as the $\delta N$ formalism or the transfer matrix method.

To track the evolution of the perturbations after Hubble radius crossing it is convenient to write the equations
in terms of the Mukhanov-Sasaki variables $Q_\sigma$ and $\delta s$ \cite{Lalak:2007vi}:
\be
\label{eq:fromLalak}
\left( \begin{array}{c} \ddot Q_\sigma \\ \ddot{\delta s} \end{array}\right)+ \left( \begin{array}{cc}3H\  & \frac{2V_{,s}}{\dot\sigma} \\
-\frac{2V_{,s}}{\dot\sigma}\  & 3H  \end{array}\right) \left( \begin{array}{c} \dot Q_\sigma \\ \dot{\delta s} \end{array}\right)
+\left[ \frac{k^2}{a^2}\mathbf{1}+   \left( \begin{array}{cc}C_{\sigma\sigma} & C_{\sigma s} \\ C_{s\sigma} & C_{ss}  \end{array}\right)  \right]
\left( \begin{array}{c}  Q_\sigma \\ \delta s \end{array}\right) = 0 \, .
\ee
The coefficients $C_{IJ}$ are given by Eqs.\eqref{eq:coo}, and can be expanded consistently
to next-order (see Appendix).
On super-Hubble scales we can neglect the $k^2/a^2$ term in Eq.(\ref{eq:fromLalak}).
Furthermore, noting that in this regime the variables $Q_\sigma$ and $\delta s$ change
slowly compared to the scale factor, we can also neglect the double time derivatives.
Thus, using Eqs.\eqref{eq:coo_sr}--\eqref{eq:css_sr} we obtain, to leading-order in the slow-roll parameters:
\begin{subequations}
\begin{align}
\label{eq:qo_hlp1}
\dot Q_\sigma/ H & = -(\eta_{\sigma\sigma}-2\varepsilon)Q_\sigma-2\eta_{\sigma s}\,\delta s \, ,\\
\label{eq:ds_hlp1}
\dot{\delta s}/H & = -\eta_{ss}\,\delta s \, .
\end{align}
\end{subequations}
Differentiating Eqs.\eqref{eq:qo_hlp1}--\eqref{eq:ds_hlp1} with respect to time and
using Eqs.\eqref{eq:sr_der1}--\eqref{eq:sr_der3}, we can obtain next-order slow-roll
expressions for $\ddot{Q}_\sigma$ and $\ddot{\delta s}$.
Plugging them back into Eq.\eqref{eq:fromLalak} and again using Eqs.\eqref{eq:coo_sr}--\eqref{eq:css_sr},
we find a \emph{next-next-order} slow-roll approximation for the full equations of motion:
\begin{subequations}
\begin{align}
\label{eq:qo_sr}
\frac{\mathrm{d}\,Q_\sigma}{\mathrm{d}N} & = A\,Q_\sigma+B\,\delta s  \, , \\
\label{eq:ds_sr}
\frac{\mathrm{d}\,\delta s}{\mathrm{d}N} & = D\,\delta s  \, ,
\end{align}
\end{subequations}
where we have defined
\begin{subequations}
\begin{align}
\label{eq:apar}
A & = -\eta_{\sigma\sigma}+2\varepsilon -\frac{1}{3}\eta_{\sigma \sigma}^2+\frac{5}{3}\varepsilon\eta_{\sigma\sigma}-\frac{4}{3}\varepsilon^2 -\frac{1}{3}\eta_{\sigma s}^2-\frac{1}{3}\xi_{\sigma\sigma\sigma}^2+\delta^{(3)}A\, ,\; \\
\label{eq:bpar}
B & = -2\eta_{\sigma s}+2\varepsilon\eta_{\sigma s}-\frac{2}{3}\eta_{\sigma s}\eta_{\sigma\sigma} -\frac{2}{3}\eta_{\sigma s}\eta_{ss}-\frac{2}{3}\xi_{\sigma\sigma s}^2+\delta^{(3)}B \, ,\; \\
\label{eq:dpar}
D & = -\eta_{ss}-\frac{1}{3}\eta_{\sigma s}^2-\frac{1}{3}\eta_{ss}^2+\frac{1}{3}\varepsilon\eta_{ss}-\frac{1}{3}\xi_{\sigma ss}^2 +\delta^{(3)}D \, ,\;
\end{align}
\end{subequations}
where $\delta^{(3)}A$, $\delta^{(3)}B$ and $\delta^{(3)}D$ denote
{\em next-to-next-order} corrections given by:
\begin{subequations}
\begin{align}
9\delta^{(3)}A  = & 28\varepsilon^3-49\varepsilon^2\eta_{\sigma\sigma} +23\varepsilon\eta_{\sigma\sigma}^2-2\eta_{\sigma\sigma}^3+15\varepsilon\eta_{\sigma s}^3-4\eta_{\sigma\sigma}\eta_{\sigma s}^2-2\eta_{\sigma s}^2\eta_{ss} + \nonumber \; \\
& +4\varepsilon\xi^2_{\sigma\sigma\sigma}-4\eta_{\sigma\sigma}\xi^2_{\sigma\sigma\sigma}-2\eta_{\sigma s}\xi^2_{\sigma\sigma s}+\Xi^3_{\sigma\sigma\sigma} 
\label{eq:sr_apar3}\, , \\
9\delta^{(3)}B  = & -58\varepsilon^2\eta_{\sigma s}+40\varepsilon\eta_{\sigma\sigma}\eta_{ss}-4\eta_{\sigma\sigma}^2\eta_{\sigma s}-4\eta_{\sigma s}^3+24\varepsilon\eta_{\sigma s}\eta_{ss}-4\eta_{\sigma\sigma}\eta_{\sigma s}\eta_{ss} - \;\nonumber \\
& - 4\eta_{\sigma s}\eta_{ss}^2-6\eta_{\sigma s}\xi^2_{\sigma\sigma\sigma}+8\varepsilon\xi^2_{\sigma\sigma s}-4\eta_{\sigma\sigma}\xi^2_{\sigma\sigma s}-4\eta_{ss} \xi^2_{\sigma\sigma s}-2\eta_{\sigma s}\xi^2_{\sigma ss} + \nonumber\\
& +2\Xi^3_{\sigma\sigma s} \, , \\
9\delta^{(3)}D = & 11\varepsilon\eta_{\sigma s}^2 -2\eta_{\sigma\sigma}\eta_{\sigma s}^2-5\varepsilon^2\eta_{ss} +2 \varepsilon\eta_{\sigma\sigma}\eta_{ss}-4\eta_{\sigma s}\eta_{ss}+5\varepsilon\eta_{ss}^2-2\eta_{ss}^3-\nonumber\\
&-6\eta_{\sigma s}\xi^2_{\sigma\sigma s}-4\eta_{ss}\xi^2_{\sigma ss}+\Xi^3_{\sigma ss} \, ,
\label{eq:sr_dpar3}
\end{align}
\end{subequations}
where $\Xi^3_{\sigma IJ}=\dot\xi^2_{\sigma IJ}/H$.
The next-order result has been first obtained in Ref.\ \cite{Peterson:2010np}.
Our inclusion of the next-next-order result allows us to estimate the accuracy of the next-order
calculation and to discuss general conditions for the applicability of the technique described here.


Eqs.\eqref{eq:qo_sr}--\eqref{eq:ds_sr} can be integrated to give
\begin{subequations}
\begin{align}
\label{eq:qo_sr_sol}
Q_\sigma(N) & = \left[ Q_{\sigma\star}+\delta s_{\star}\int_{N_{\star}}^N \mathrm{d}N''B(N'')e^{\int_{N_{\star}}^{N''}\mathrm{d}N'(D(N')-A(N'))}\right]
e^{\int_{N_{\star}}^N\mathrm{d}\tilde N\,A(\tilde N)} \, , \\
\label{eq:ds_sr_sol}
\delta s(N) & = \delta s_{\star} e^{\int_{N_{\star}}^N \mathrm{d}\tilde N\,D(\tilde N)} \, , \;
\end{align}
\end{subequations}
where $Q_{\sigma\star}$ and $\delta s_{\star}$ are the initial conditions at $N_{\star}$.
Taking $N_{\star}$ to be the e-fold at which the mode of interest crossed the Hubble radius
and applying Eqs.\eqref{eq:curv_spectrum}--\eqref{eq:isocurv_spectrum}, we find that the instantaneous
power spectra for ${\mathcal{R}}$ and $\mathcal{S}$ can be written as
\begin{subequations}
\begin{align}
\label{eq:ps_qo_sr_sol}
\mathcal{P}_{\mathcal{R}} & = \mathcal{P}_{\star} \left[ 1+\left(\int_{N_{\star}}^N \mathrm{d}N''B(N'')e^{\int_{N_{\star}}^{N''}\mathrm{d}N'G(N')}\right)^2 +c_\star \int_{N_{\star}}^N \mathrm{d}N''B(N'')e^{\int_{N_{\star}}^{N''}\mathrm{d}N'G(N')}\right] \ \; \\
\label{eq:ps_ds_sr_sol}
\mathcal{P}_{\mathcal{S}} & = \mathcal{P}_{\mathcal{S}\star} e^{2\int_{N_{\star}}^N\mathrm{d}N'\,G(N')} \, , \;
\end{align}
\end{subequations}
where $G=D-A$, $\mathcal{P}_{\star}=H_{\star}^2/8\pi^2\varepsilon_{\star}$ and
$\mathcal{P}_{\mathcal{S}\star}$ is given by Eq.\eqref{eq:isocurv_spectrum} setting $k_{\star}=k$
and $-k\tau=1$.
The relative correlation between curvature and isocurvature perturbations at
Hubble radius crossing is denoted by $c_\star=-2\eta_{\sigma s\star}(2-\ln 2-\EulerGamma)$ \cite{Byrnes:2006fr}.
In the following section we will apply the solutions \eqref{eq:ps_qo_sr_sol}--\eqref{eq:ps_ds_sr_sol}
to specific inflationary models.
We will compare the accuracy of these solutions to that found by numerically integrating
the full equations of motion for the perturbations.

Before we proceed to the numerical results, we summarize the
qualitative features of the solutions described here with typical time evolution of the
slow-roll parameters found in \S\ref{sec:model}. 
Using Eq. (\ref{eq:sr_dersol}) and $\eta_{ss}^{(0)}\approx-0.1$, the isocurvature perturbations grow steadily after the Hubble radius
crossing and later source the curvature perturbations for $\mathcal{O}(1/|\eta_{ss}^{(0)}|)$ e-folds.
In this case, the need for next-next-order contributions in Eqs. 
(\ref{eq:ps_qo_sr_sol})--(\ref{eq:ps_ds_sr_sol})
comes from the fact that the expression $-2\eta_{ss}^2\Delta N/3$ can become a large fraction of $\eta_{ss}$ if there are $\mathcal{O}(1/|\eta_{ss}^{(0)}|)$ of e-folds between the Hubble radius crossing and the time when the coupling between the curvature and iscorvature perturbations reaches its maximal value. 
It also follows from Eqs. (\ref{eq:sr_apar3})--(\ref{eq:sr_dpar3})
that next-to-next order corrections should be negligible in this case.

A different situation is encountered for the pattern described in 
Eq. (\ref{eq:sr_dersol2}). Here the isocurvature perturbations
slowly decay due to nonzero $2\varepsilon-\eta_{\sigma\sigma}$ until a spike in $\eta_{\sigma\sigma}$
produces a short increase in $\mathcal{P}_\mathcal{S}$ which, as $\eta_{\sigma s}$ is simultaneously
driven to large values, can give a sizable sourcing of the curvature perturbations. This transient
growth of the slow-roll parameters makes the use of the next order (or, generically, 
higher order) calculation necessary; whether this
accuracy is sufficient depends on how large the slow-roll parameters actually become. As can be seen
from Eqs. (\ref{eq:sr_apar3})--(\ref{eq:sr_dpar3}), the terms involving $\varepsilon$ come with large numerical
coefficients, so when $\varepsilon$ becomes too close to 1, the splitting between different orders
in slow-roll makes no sense and the perturbative calculation described here breaks down.

\section{Numerical examples}
\label{sec:numerics}

Hybrid inflation \cite{Linde:1991km} is unquestionably the most
	popular inflationary model with
	two scalar fields. In its original version,
	the inflaton slowly rolls down the potential until it reaches a critical
	value. Then, the waterfall field becomes tachyonic and
	quickly rolls down to the minimum of the potential, thereby terminating inflation.
	Such dynamics leads, however, to a blue spectral index, inconsistent with
	observations. This motivated a number of authors
	\cite{Clesse:2010iz,Abolhasani:2010kn,Kodama:2011vs}
	to consider variants
	of hybrid inflation admitting a much milder transition between the inflaton-dominated
	and waterfall-dominated regimes of inflation\footnote{See \cite{Martin:2011ib} for a different approach where the authors investigate the importance of quantum diffusion on the transition.}. In these models, the waterfall field remains
	tachyonic, but the magnitude of its mass parameter can be much smaller than the Hubble
	scale.
	In this case one has to analyze carefully the dynamics of the coupled system
	describing the behaviour of the perturbations of
	the two scalar fields. In \S\ref{sec:num_hyb_intro} we investigate
	precisely these types of hybrid inflation models, whereas \S\ref{sec:num_2}
	will be devoted to models where the inflationary trajectory
	has a turn in field space during which some slow-roll parameters may become
	close to one while maintaining the slow-roll condition $\varepsilon\ll1$.

\subsection{A case study of hybrid inflation}
\label{sec:num_hyb_intro}

	As a concrete example, we shall study one particular inflationary trajectory
	which was also investigated in Ref.\ \cite{Kodama:2011vs}. The potential of the model is
	\be
	V(\varphi,\chi) = \Lambda^4 \left[ \left( 1-\frac{\chi^2}{v^2}\right)^2+\frac{\varphi^2}{\mu^2} +\frac{2\varphi^2\chi^2}{\varphi_c^2v^2}\right]\, ,
	\ee
	where we adopt their notation and $v=0.10$, $\varphi_c=0.01$ and $\mu=1.00\times 10^3$. 
The overall normalization of the potential $\Lambda$ does not play any role in this discussion and can be arbitrarily chosen
	to ensure the correct normalization of the curvature perturbations \cite{Komatsu:2010fb}.
	Our particular trajectory corresponds to the one shown in Fig.~4 of Ref.~\cite{Kodama:2011vs}:
	it starts at $\varphi_0=1.00\times 10^{-2}$ and $\chi_0=1.63\times 10^{-9}$,
	producing 62 e-folds of inflation. In the following, we shall mainly focus on the
	evolution of modes leaving the Hubble radius 8 e-folds after this initial time,
	so it is convenient to associate the beginning of the zeroth e-fold with that moment
	(negative number of e-folds will then correspond to earlier times).
	
\begin{figure}
\begin{center}
\includegraphics*[height=5.6cm]{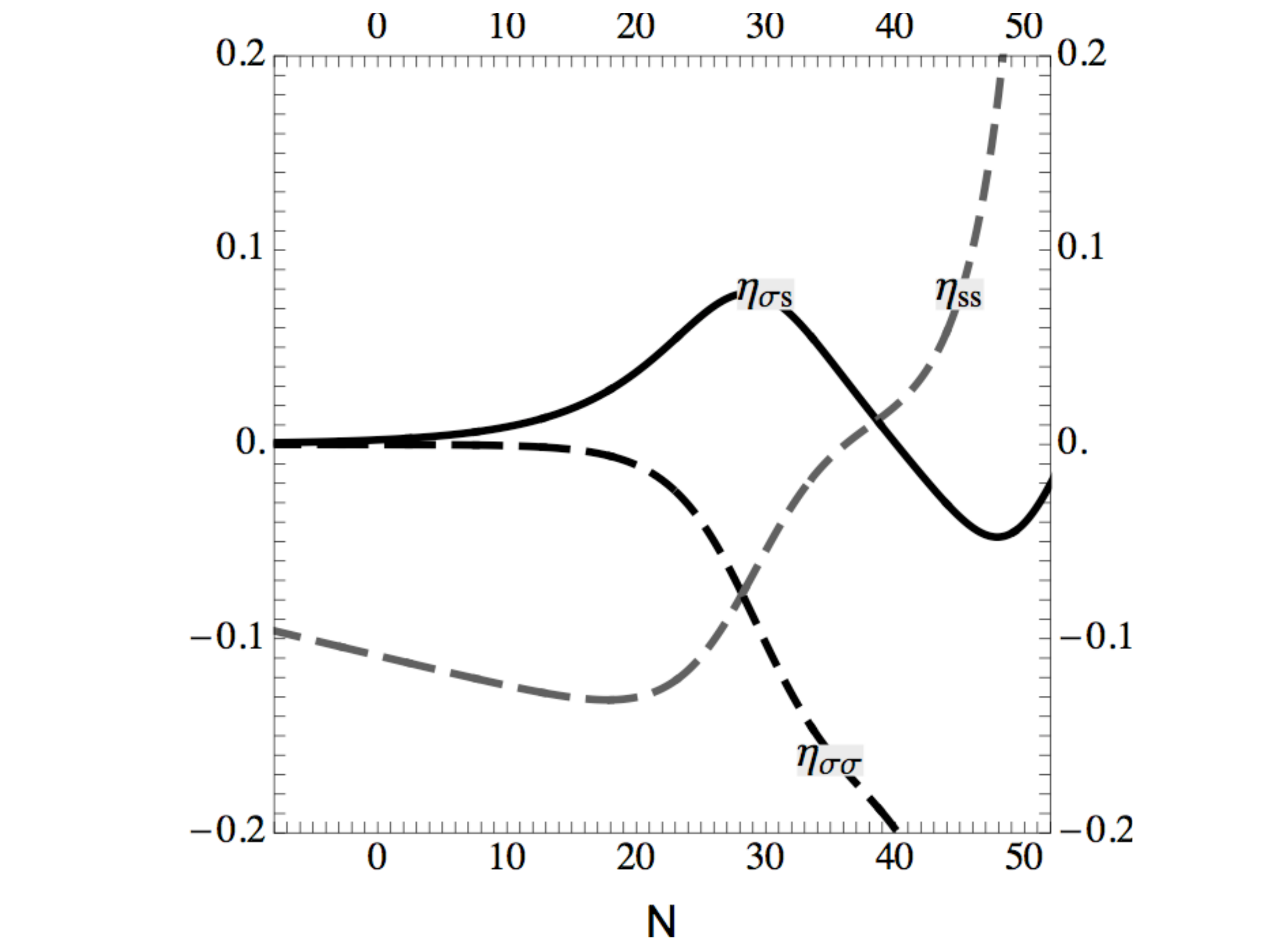}
\hspace{0.2cm}
\includegraphics*[height=5.6cm]{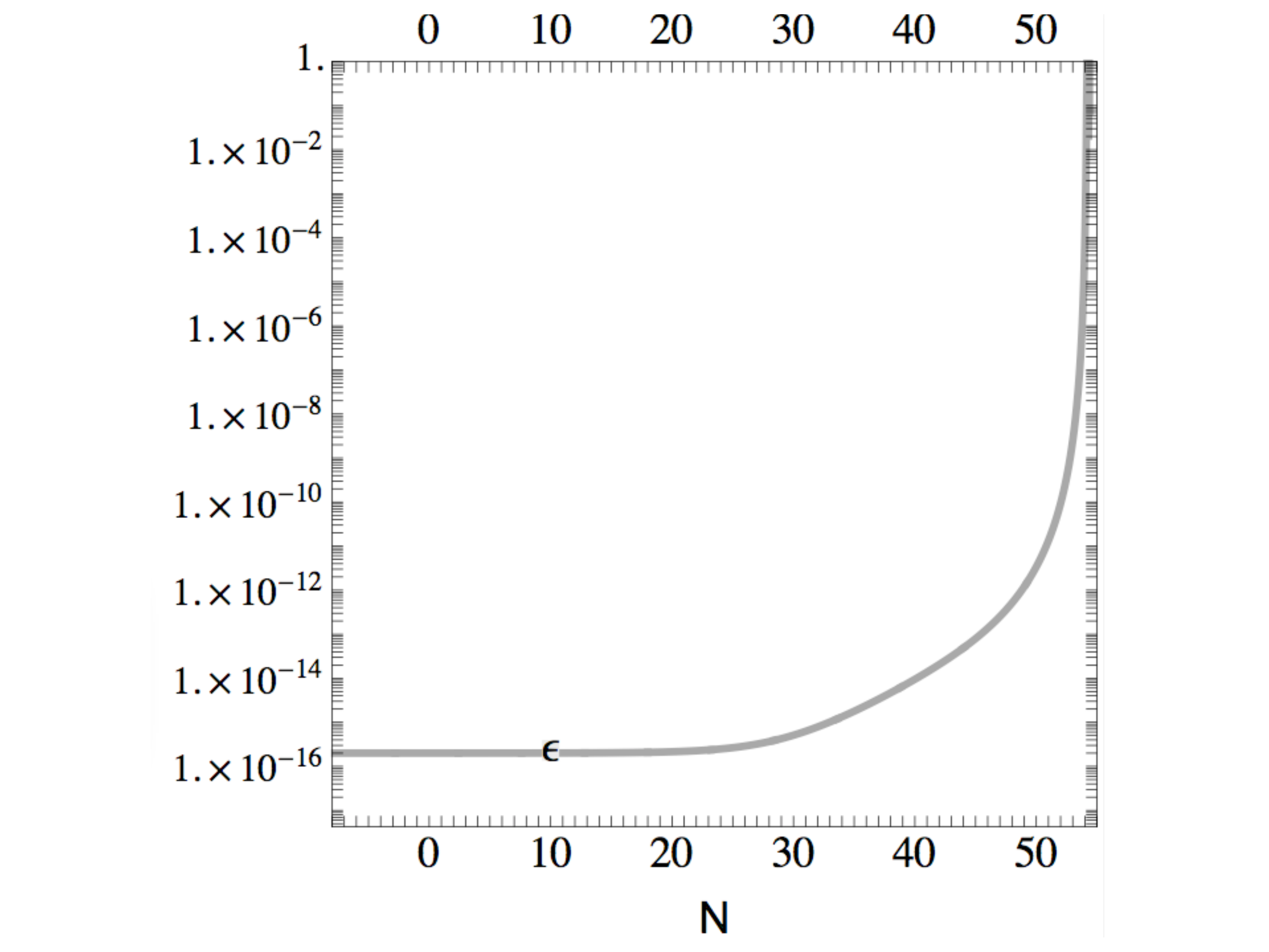}
\end{center}
\caption{The evolution of the slow-roll parameters	$\varepsilon$, $\eta_{\sigma\sigma}$, $\eta_{\sigma s}$ and $\eta_{ss}$
	for the inflationary trajectory introduced in \S \ref{sec:num_hyb_intro}. \label{f:sr_par}}
\end{figure}

	In Fig.\ \ref{f:sr_par} we show the time evolution of the slow-roll parameters
	$\varepsilon$, $\eta_{\sigma\sigma}$, $\eta_{\sigma s}$ and $\eta_{ss}$.
	It is clear that, for the initial $30$ or so e-folds, the hierarchy
	$\left( \varepsilon,\,|\eta_{\sigma\sigma}|,\,|\eta_{\sigma s}| \right) \ll|\eta_{ss}|<1$ holds, realizing
one of the possibilities discussed in \S\ref{sec:sr_diag}. 
We note that the evolution
of the slow-roll parameters $\eta_{\sigma\sigma}$, $\eta_{\sigma s}$ and $\eta_{ss}$
qualitatively agrees with the approximate solution 
(\ref{eq:sr_dersol}).

\begin{figure}
\begin{center}
\includegraphics*[height=6.5cm]{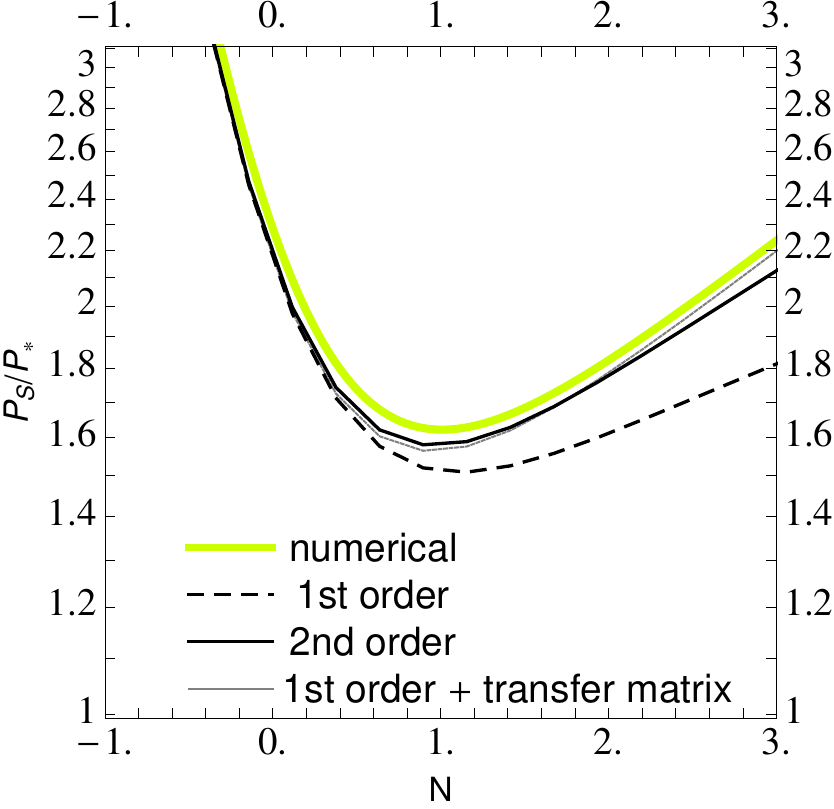}
\hspace{0.2cm}
\includegraphics*[height=6.5cm]{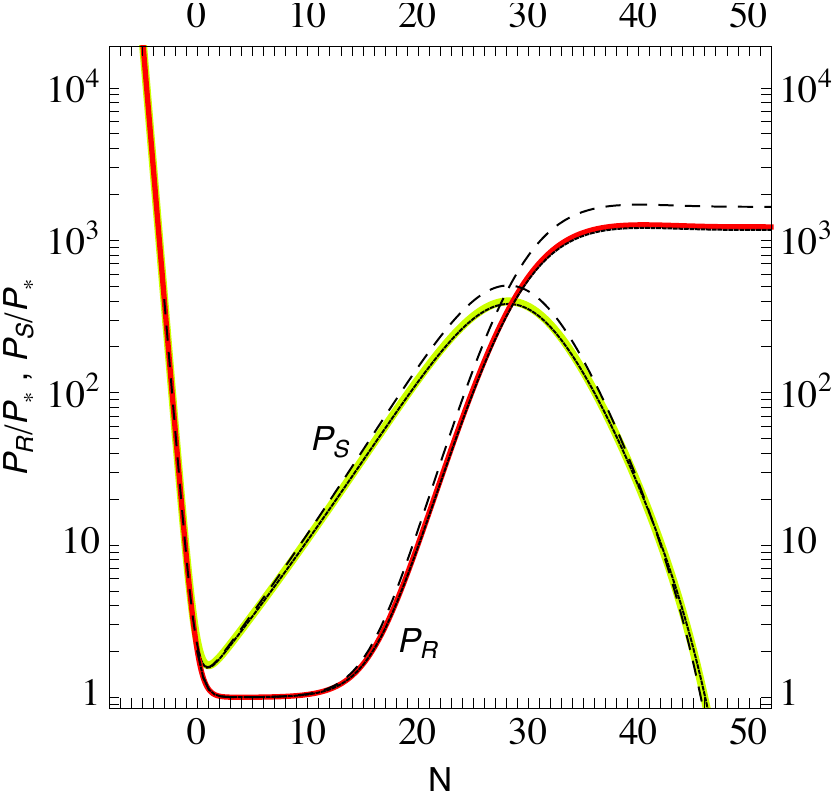}
\end{center}
\caption{Left panel: the power spectrum of the curvature perturbations, $\mathcal{P}_{\mathcal{S}}$,
normalized to the single-field power spectrum of curvature perturbations $\mathcal{P}_{\star}$
calculated in four different ways as described in the text. Right panel:
power spectra of the curvature and isocurvature perturbations
($\mathcal{P}_\mathcal{R}$ and $\mathcal{P}_\mathcal{S}$, respectively)
for the case study of hybrid inflation described in \S\ref{sec:num_hyb_intro}.
They were calculated numerically (red and yellow lines) or taken from analytical solutions
(\ref{eq:ps_qo_sr_sol})--(\ref{eq:ps_ds_sr_sol}) with parameters $B$ and $G$
expanded to leading (black dashed lines) or next-order (black solid lines)
in the slow-roll parameters. The latter practically overlap with numerical results.
\label{f:ps_hc}}
\label{f:ps_after}
\end{figure}

We observe from Fig.~\ref{f:ps_hc} that the $\eta_{ss\star}^2$ corrections in
Eq.\eqref{eq:isocurv_spectrum} are important for ensuring the accuracy of the
result for a few e-folds.  The results are normalized to the value
$\mathcal{P}_{\star}=H_{\star}^2/8\pi^2\varepsilon_{\star}$ of the single-field
power spectrum of curvature perturbations.  In the left panel, the evolution of
$\mathcal{P}_\mathcal{S}$ is calculated in four different ways.  The yellow
line corresponds to the numerical integration of the full equations of motion.
The dashed black line is the prediction of Eq.\eqref{eq:isocurv_spectrum} neglecting $\eta_{ss\star}^2$,
while the solid black line corresponds to the full expression taking into account next-order effects.
Finally, the thin grey solid line is the result of applying the transfer matrix formalism discussed in \S\ref{sec:2pf}
with a leading-order slow-roll result for the power spectrum taken as the initial condition at $N=0$.
 In these calculations, we have identified the number of e-folds $N$ with $\ln(-k_\star\tau)$,
dropping the $\mathcal{O}(\varepsilon_\star)$ correction in Eq.(\ref{eq:sca3}). Since $\varepsilon$ is
very small until the end of inflation (cf.\ Fig.\ \ref{f:sr_par}), the error introduced by such an approximation is negligible.
These results show that, for $|\eta_{ss\star}|=\Or(10^{-1})$, one must take into account
corrections quadratic in $\eta_{ss}$ in order to obtain a faithful estimate of the power
spectrum of the isocurvature perturbations, valid for a
couple of e-folds after Hubble radius crossing.

Adhering to the particular trajectory in the model of hybrid inflation considered here,
we have calculated the evolution of
curvature and isocurvature perturbations during inflation for one particular mode leaving
the Hubble radius at $N=0$. The results are shown in Fig.\ \ref{f:ps_after} (right panel).
Yellow and red lines correspond to the curvature and isocurvature perturbations, respectively,
whose evolution was calculated numerically from the full set of equations of motion
(\ref{eq:eom_usigma})--(\ref{eq:eom_us}).
Black dashed (solid) lines show the predictions of our analytical solutions (\ref{eq:ps_qo_sr_sol})--(\ref{eq:ps_ds_sr_sol})
with coefficients $B$ and $G$ expanded to next to leading order in the slow-roll parameters.

We find good agreement between the full numerical solution
and the approximate solutions (\ref{eq:ps_qo_sr_sol})--(\ref{eq:ps_ds_sr_sol}) obtained by including
next-order slow-roll corrections.
On the other hand, terminating the expansion at leading-order in slow-roll parameters gives inaccurate results.
This behavior is a direct consequence of the fact that with $\eta_{ss}\simeq -0.1$ the
isocurvature perturbation is temporarily tachyonic, and grows for around twenty-five e-folds or so
before the coupling between the curvature and the isocurvature perturbations becomes
large, and the curvature perturbations are sourced by the amplified isocurvature modes.
A correction of $-2\eta_{ss}^2\Delta N/3$ in the exponent of
(\ref{eq:ps_ds_sr_sol})
can then easily be as large as around $20\%$, which is the magnitude of the effect we observe.

In the $\delta N$ formalism, the curvature perturbations on large scales are given by
\be
\mathcal{R}(t_f,\mathbf{x}) = \mathcal{N}(t_f,t_{\star},\mathbf{x})- N(t_f,t_{\star})\ ,
\ee
where $\mathcal{N}(t_f,t_{\star},\mathbf{x})$ is the number of e-folds between an initial
flat hypersurface at time $t_{\star}$ and a uniform density hypersurface at
time $t_f$, and $N(t_f,t_{\star})=\int_{t_{\star}}^{t_f}H(t)\,\mathrm{d}t$.
Following this definition, in two-field inflationary models
one can write the power spectrum of the curvature
perturbations as
\be
\label{eq:dn_res}
\mathcal{P}_{\mathcal{R}} = \mathcal{P}_{Q_\varphi} \left(\frac{\partial N}{\partial\varphi}\right)^2
+\mathcal{P}_{Q_\chi} \left(\frac{\partial N}{\partial\chi}\right)^2 \, ,
\ee
where $Q_\varphi$ and $Q_\chi$ are the Mukhanov-Sasaki variables associated with the
perturbations of the scalar fields $\varphi$ and $\chi$, respectively.
Also, $\mathcal{P}_{Q_\varphi}$
and $\mathcal{P}_{Q_\chi}$ are the power spectra evaluated at $t_{\star}$ and the partial derivatives
are taken with respect to the values of the fields at $t_{\star}$. 
Eq.\ (\ref{eq:dn_res}) does not rely on the slow-roll approximation.
Usually, one further simplifies this
result, associating $t_{\star}$ with a time soon after Hubble radius crossing and
taking $\mathcal{P}_{Q_\varphi}=\mathcal{P}_{Q_\chi}=H_{\star}^2/4\pi^2$.
However, for the inflationary trajectory considered here the latter assumption
is not accurate enough: at Hubble radius crossing the power spectrum of the
isocurvature perturbations is enhanced by nearly $20\%$, and this effect must be included in the
calculation of the final power spectrum of the curvature perturbations.
Upon identifying $Q_\chi$ with $\delta s$, one notes that
this correction is easily recognized as the $\mathcal{O}(\eta_{ss}^2)$
term in Eq.(\ref{eq:isocurv_spectrum}).
Importantly, this term should not be interpreted as a next-order correction
to the $\delta N$ result, since the latter is already fully non-linear in the slow-roll expansion; instead
it corresponds to 
the next-order correction to the initial conditions in the $\delta N$ formalism.

To close this section we comment on how some of the subtleties discussed above impact existing results
found in the literature. First, we note that $Q_\sigma$ and $\delta s$ are
two independent quantum fields. In order to obtain their power spectra, one should
solve the equations of motion for the wave functions twice, assuming either $\delta s=0$
or $Q_\sigma=0$ as initial conditions; the two results should be then added in
quadratures. Ref.\  \cite{Abolhasani:2010kn} solves the equations of motion for
the wave functions only once, which leads to a spurious dip in the plot of $|Q_\sigma|$
shown in logarithmic scale. However, thanks to a fortunate coincidence,
the final result is numerically correct;  the final curvature perturbations are dominated
by the contribution of the isocurvature perturbations,
so the initial curvature perturbation can be safely neglected.
In Ref.\ \cite{Kodama:2011vs} the sourcing of the curvature
perturbations by the isocurvature perturbations was neglected, which not only leads to
a final amplitude of the curvature perturbations that is orders of magnitude smaller than
the full result, but also gives an incorrect estimate $n_s-1=-6\varepsilon_{\star}+2\eta_{\sigma\sigma\star}$
for the scalar spectral index. 
It follows from (\ref{eq:ps_qo_sr_sol}) that \cite{Byrnes:2006fr}
\begin{equation}
n_s-1\simeq -(6-4c^2)\varepsilon_\star+2(1-c^2)\eta_{\sigma\sigma\star}+2c^2\eta_{ss\star} \, ,
\end{equation}
where $c^2=1-\mathcal{P}_\star/\mathcal{P}_\mathcal{R}$ 
is the isocurvature-sourced fraction of the curvature perturbations
(here we assumed $c_\star=0$).
However, for the trajectory discussed here $c^2\sim 0.999$, so $n_s\approx 0.8$,
which is 
excluded by WMAP results
\cite{Komatsu:2010fb}
(irrespective of the constraints from the normalization of the perturbations \cite{Abolhasani:2011yp}). 
More generally, with negligible $\varepsilon_\star$, $\eta_{\sigma\sigma\star}$ and $\eta_{\sigma s\star}$,
we find that 
$c^2=(n_s-1)/2\eta_{ss\star}$, so with $\eta_{ss\star}\sim-0.1$ and $n_s-1\sim-0.04$ we can only have $c^2\sim0.2$. 
Fig.~\ref{f:ps_hc} shows that, compared to the leading-order results, for $\eta_{ss\star}\sim-0.1$ we have $\sim10\%$ correction 
to $c^2$, coming from a similar correction to
$\mathcal{P}_{\mathcal{S}\star}$.
From this estimate we can conclude that for phenomenologically viable models
with the pattern of the slow-roll parameters considered here, the next-order corrections
at Hubble radius crossing result in corrections to the power spectrum of the curvature
perturbations at the level of a few percent. This is comparable to the expected accuracy
of the {\em Planck} measurements.

\subsection{A case study of double quadratic inflation}
\label{sec:num_2}

Double quadratic inflation models \cite{Langlois:1999dw}, given by the potential
\be
V(\varphi,\chi) = \frac{1}{2}m_\varphi^2 \varphi^2+\frac{1}{2}m_\chi^2\chi^2 \, ,
\ee
are a well-studied class of multi-field inflation models.
Here we study a trajectory
closely resembling the example investigated in Ref.~\cite{Lalak:2007vi}:
we take $m_\chi/m_\varphi=7$ and start the evolution of the homogeneous fields at
$\varphi_0=8$ and $\chi_0=8$ (the overall normalization of the potential only affects the
normalization of the perturbations).
Inflation is initially driven by $\chi$. Then, around $N=10$
the field space trajectory rapidly changes direction and there is a slight deviation
from slow-roll, as measured by the $\varepsilon$ parameter\footnote{In the models considered
here and
in \S\ref{sec:num_4}, the number of e-folds is smaller than required by observations. However,
these models are only intended as illustrations of the general effects we are studying.}.

\begin{figure}
\begin{center}
\includegraphics*[height=6.5cm]{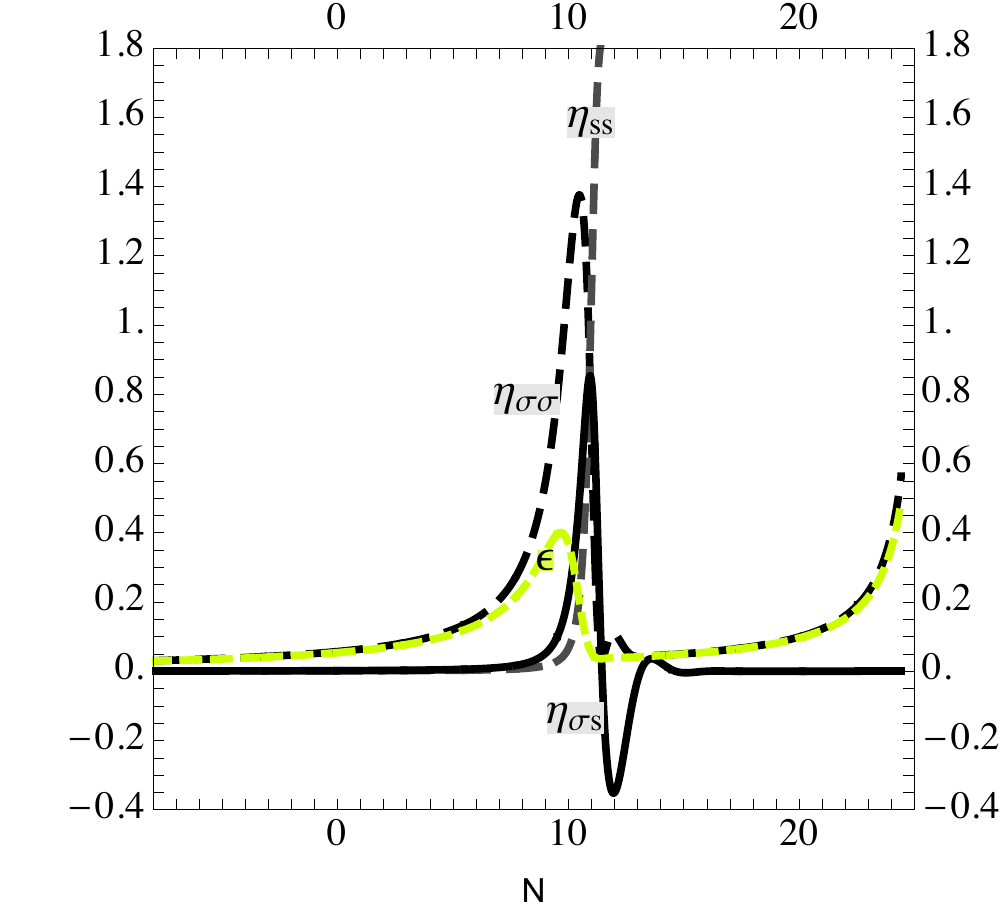}
\hspace{0.2cm}
\includegraphics*[height=6.5cm]{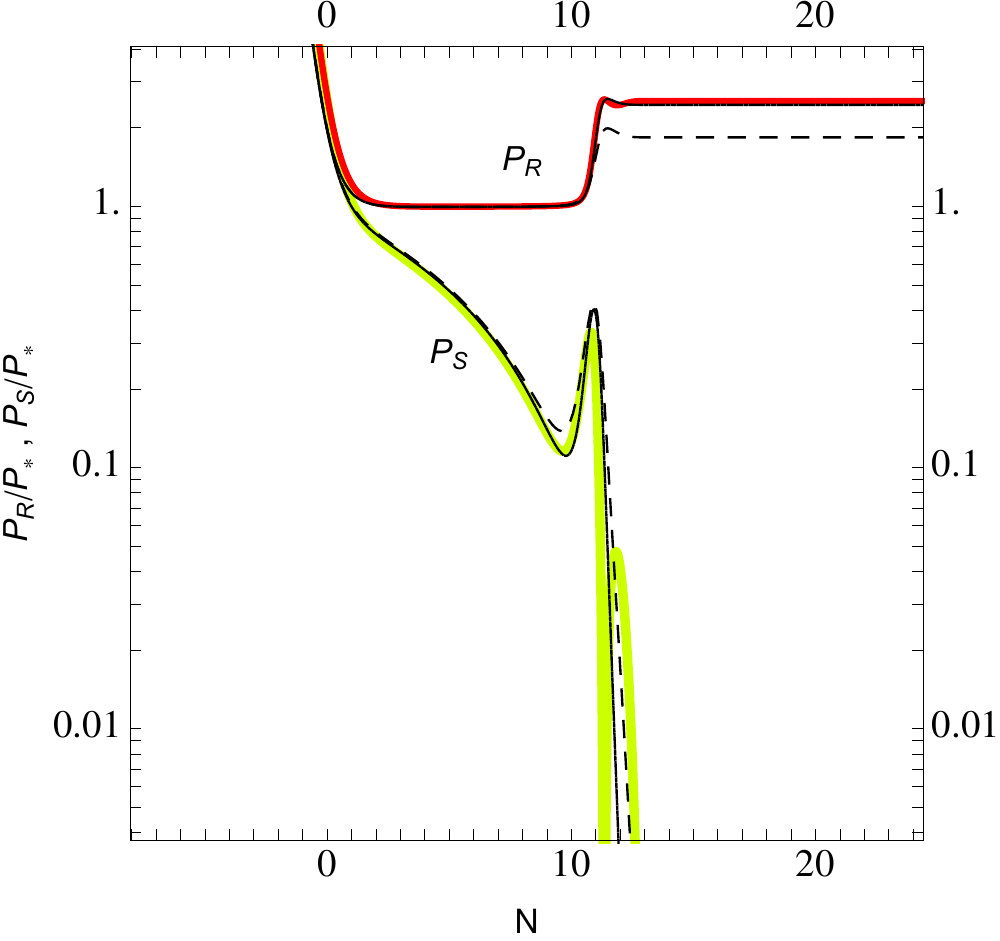}
\end{center}
\caption{
Case study of double quadratic inflation described in Sec. \ref{sec:num_2}.
The left panel shows the evolution of the slow-roll parameters
	$\varepsilon$, $\eta_{\sigma\sigma}$, $\eta_{\sigma s}$ and $\eta_{ss}$.
The right panel shows the
power spectra of the curvature and isocurvature perturbations
($\mathcal{P}_\mathcal{R}$ and $\mathcal{P}_\mathcal{S}$, respectively)
calculated numerically (red and yellow lines) or taken from the analytical solutions
(\ref{eq:ps_qo_sr_sol})--(\ref{eq:ps_ds_sr_sol}) with parameters $B$ and $G$
expanded to leading (black dashed lines) or next-order (black solid lines)
in the slow-roll parameters.
\label{f:ps_after2}}
\end{figure}

The evolution of the slow-roll parameters
$\varepsilon$, $\eta_{\sigma\sigma}$, $\eta_{\sigma s}$ and $\eta_{ss}$
is shown in Fig.~\ref{f:ps_after2}, where we also display the evolution
of the perturbations. Since in this example $\xi_{IJK}^2=0$, the coupling $B$
between the curvature and isocurvature perturbations on super-Hubble scales,
given by Eq.(\ref{eq:bpar}), is proportional to $\eta_{\sigma s}$,
which explains a small decrease in $\mathcal{P}_\mathcal{R}$
after reaching the maximum at $N\simeq11$.
The slow-roll expansion becomes unreliable when $\eta_{ss}$ becomes much larger than one.
Our next-order slow-roll expansion is, however, quite reliable before that happens.
The discrepancy between the results calculated up to leading and next-order in slow-roll
comes from the contributions to the parameter $B$, with the next-order contribution only
twice smaller than the first order one.
At later times the isocurvature perturbations become very massive and decay rapidly,
no longer affecting the evolution of the curvature modes.
The most striking result of the breakdown of the slow-roll expansion is the
oscillatory feature in the power spectrum of the
isocurvature perturbations $\mathcal{P}_\mathcal{S}$,
resulting from a sudden change of the mass parameter $3\eta_{ss}H^2$ of this perturbation.
This also explains an interesting feature in Fig.~\ref{f:ps_after2}: after decaying for a short period (when $N\simeq11$) the quantity $\mathcal{P}_{\mathcal{R}}$ increases slightly. This effect goes beyond our analytic approximation, but has almost no impact on the final value of the power spectrum.

\subsection{A case study of double quartic inflation}
\label{sec:num_4}

This example originates from the model of matrix inflation \cite{Ashoorioon:2009sr}.
With certain simplifying assumptions the potential of this model can be written as
\be
V(\varphi,\chi)=\frac{1}{4}\lambda_\varphi\varphi^4+ \frac{1}{4}\lambda_\chi\chi^4 \, .
\ee
We investigated the second of the exemplary inflationary trajectories
studied in \cite{Ashoorioon:2009sr}, taking
$\lambda_\varphi/\lambda_\chi=410$ with $\varphi_0=11.2$ and $\chi_0=9.1$.
Initially, inflation is driven by $\varphi$, then, around $N=8$,
the trajectory in the fields space rapidly changes direction and there is a deviation
from slow-roll, as measured by the $\varepsilon$ parameter.

\begin{figure}
\begin{center}
\includegraphics*[height=6.5cm]{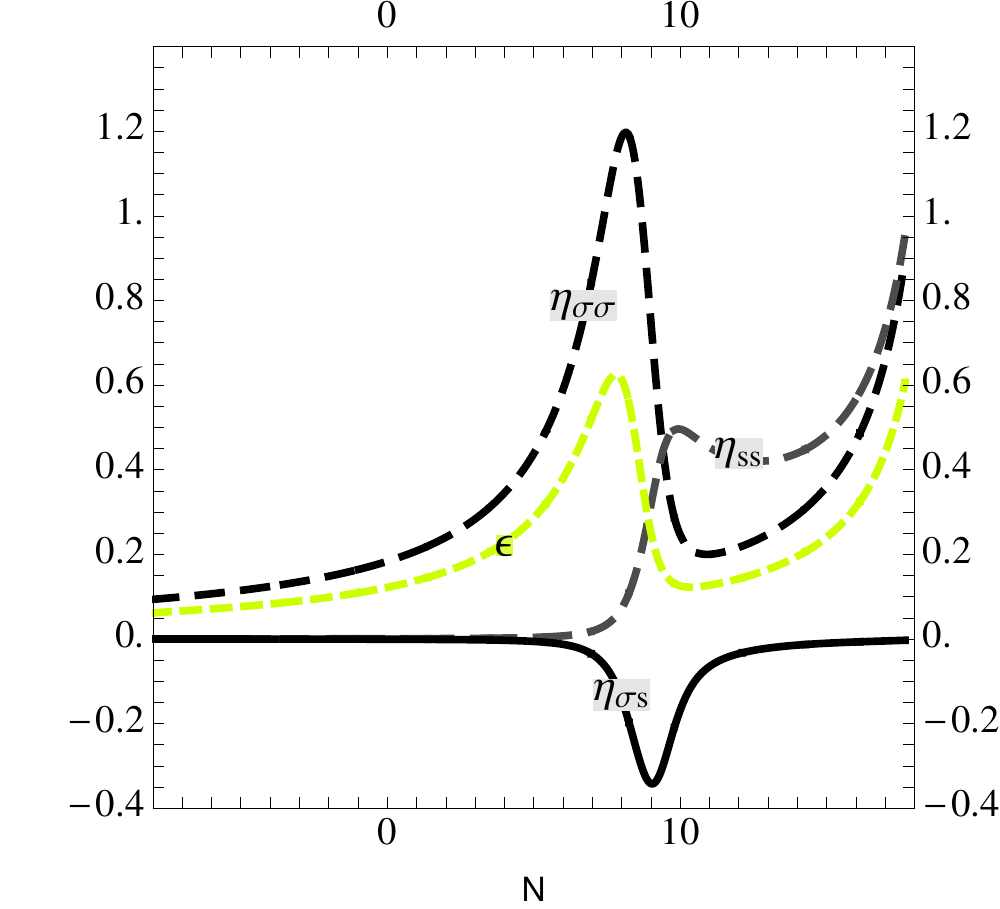}
\hspace{0.2cm}
\includegraphics*[height=6.5cm]{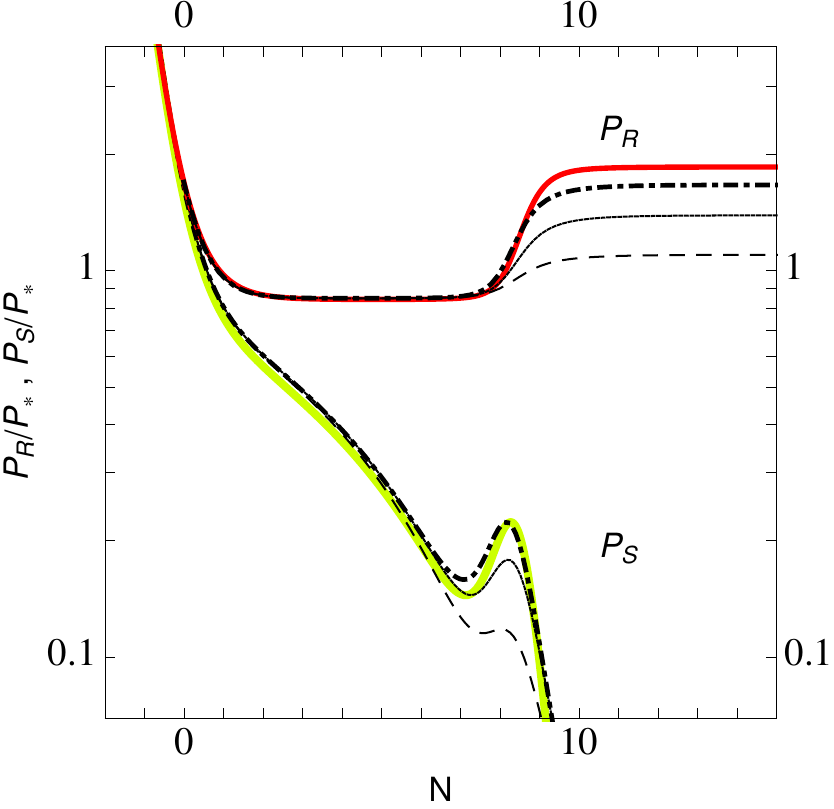}
\end{center}
\caption{
Case study double quartic inflation described in Sec.\ \ref{sec:num_4}.
Left panel shows the evolution of the slow-roll parameters
	$\varepsilon$, $\eta_{\sigma\sigma}$, $\eta_{\sigma s}$ and $\eta_{ss}$.
	Right panel shows the
power spectra of the curvature and isocurvature perturbations 
($\mathcal{P}_\mathcal{R}$ and $\mathcal{P}_\mathcal{S}$, respectively)
calculated numerically (red and yellow lines) or taken from analytical solutions 
(\ref{eq:ps_qo_sr_sol})-(\ref{eq:ps_ds_sr_sol}) with parameters $B$ and $G$
expanded to the leading order (black dashed lines), next order (black solid lines)
or next-to-next order (black dash-dotted lines) 
in the slow-roll parameters.
\label{f:ps_after3}}
\end{figure}

The evolution of the slow-roll parameters
	$\varepsilon$, $\eta_{\sigma\sigma}$, $\eta_{\sigma s}$ and $\eta_{ss}$
is shown in Fig.\ \ref{f:ps_after3}, where we also display the evolution
of the perturbations. It is a clear example of the limitation of the method
presented in \S\ref{sec:2pfb}:
the higher order terms in Eqs. (\ref{eq:sr_apar3})--(\ref{eq:sr_dpar3})
contribute 
terms with coefficients much larger than one and containing powers of $\varepsilon$
in Eqs. (\ref{eq:ps_qo_sr_sol})--(\ref{eq:ps_ds_sr_sol}).
Hence whenever $\varepsilon$ becomes too close to 1 during inflation, the expansion
in the power series of the slow-roll parameters starts to break down.
The example shown here is on the verge of applicability of the perturbative expansion
in the powers of the slow-roll parameters, and it can be seen from Fig.\ \ref{f:ps_after3}
that the next-to-next order provides a fairly improved estimate of the
curvature and the isocurvature perturbations.

\section{Conclusions and outlook}
\label{sec:conclusions}

In the present era of precision cosmology, the \textit{Planck} satellite should deliver high-precision data in the near future. Such observational standards need to be followed by equally good theoretical precision so that one can make the most efficient use of the \textit{Planck} data. Whilst the inflationary power spectrum at lowest-order in slow-roll can provide an accurate description of cosmological evolution in many scenarios, there exist models where one requires slow-roll effects beyond lowest-order for an accurate description of the evolution of the perturbations.

In this paper we have studied the evolution of the curvature and isocurvature perturbations in two-field inflation models to next-next-order in the appropriate slow-roll parameters. Using popular examples of two-field inflationary models, a variant of hybrid inflation and double quadratic/quartic inflation, we have found that calculating the two-point correlations of the isocurvature modes at next-order 
in $\eta_{ss}$  provides a very precise estimate of the initial amplitude of the perturbations \textit{around} the Hubble radius. The isocurvature perturbations can source the curvature perturbations during a turn in the inflationary trajectory in field space: all slow-roll parameters can (temporarily) increase at this instance, which
justifies keeping next-order effects in the analysis. To follow the dynamical evolution, we have applied the transfer matrix method and have found a remarkable agreement with the explicit numerical integration of the equations of motion.

In conclusion, our results
show that keeping slow-roll corrections to higher order is necessary to obtain an
accurate description of perturbations on super-Hubble scales.  We reached this
conclusion using knowledge of a single inflationary trajectory, which had the advantage
of making the precision of our semi-analytical calculations comparable to those of the
full numerical methods.   In the examples we considered we found that this enables this
class of models to be predictive at an accuracy required for comparison with data
expected from \textit{Planck}.

\acknowledgments
We would like to thank Amjad Ashoorioon, Zygmunt Lalak, 
David Seery and Gary Shiu for discussions, 
and Courtney Peterson,  Max Tegmark and Masahide Yamaguchi for correspondence.
SC and SW would like to thank Robert Brandenberger and McGill University for hospitality and financial support.
SC thanks KITP for hospitality during the final stages of this work.
SW thanks DAMTP, Cambridge University and Cornell University for hospitality. 

\vspace*{0.2cm}
\noindent AA is supported by a CTC Fellowship at DAMTP, Cambridge University; SC 
is supported by the Cambridge-Mitchell Collaboration in Theoretical
Cosmology, and the Mitchell Family Foundation.
RHR is supported by Funda\c{c}\~{a}o para a Ci\^{e}ncia e a Tecnologia
(Portugal) through the grant SFRH/BD/35984/2007.  KT is partly supported
by the MNiSW grant N N202 091839. 
SW is supported by the Syracuse University College of Arts and Sciences.
This work is supported in part by STFC, UK.

\appendix
\section{Equations of motion for the perturbations}

Some of the results reviewed in \S\ref{sec:model} and \S\ref{sec:2pf}
require rather lengthy expressions which might obscure the main line of thought.
For completeness of the discussion, we have collected them here.

The coefficients $C_{IJ}$
          in the equations of motion \eqref{eq:eom_usigma} and \eqref{eq:eom_us} are:
			\begin{subequations}
			\label{eq:coo}
	\begin{align}
		C_{\sigma\sigma} & = V_{,\sigma\sigma} - \left(\dfrac{V_{,s}}{\dot{\sigma}}\right)^2
		+2 \dot{\sigma} \dfrac{V_{,\sigma}}{H} +3\dot{\sigma}^2 -
		\dfrac{\dot{\sigma}^4}{2H^2} \, ,
					\\
			C_{\sigma s} & = 6H \dfrac{V_{,s}}{\dot{\sigma}} +2 \dfrac{V_{,\sigma} V_{,s}}{\dot{\sigma}^2}
			+2V_{\sigma s} +\dot{\sigma} \dfrac{V_{,s}}{H}  \, ,
					\\
		C_{s\sigma} & =  -6H \dfrac{V_{,s}}{\dot{\sigma}} -
			2 \dfrac{V_{,\sigma} V_{,s}}{\dot{\sigma}^2} +
			\dot{\sigma} \dfrac{V_{,s}}{H}  \, ,
					\\
			C_{ss} & = V_{,ss} - \left(\dfrac{V_{,s}}{\dot{\sigma}}\right)^2  \, ,
			\end{align}
		\end{subequations}
which are exact to all orders in slow-roll.
     We can expand these coefficients
          uniformly to next-order in the slow-roll parameters:
			\begin{subequations}
	\begin{align}
	\label{eq:coo_sr}
		C_{\sigma\sigma} & \simeq 3H^2\left( \eta_{\sigma\sigma}-2\varepsilon-\frac{1}{3}\eta_{\sigma s}^2+\frac{4}{3}\varepsilon\eta_{\sigma\sigma}-\frac{2}{3}\varepsilon^2 \right)  \; ,
					\\
			C_{\sigma s} & \simeq 3H^2\left( 2\eta_{\sigma s}+\frac{2}{3}\eta_{\sigma s}\eta_{\sigma\sigma} \right)  \; ,
					\\
		C_{s\sigma} & \simeq 3H^2\left( -\frac{2}{3}\eta_{\sigma s}\eta_{\sigma\sigma} +\frac{4}{3}\varepsilon\eta_{\sigma s} \right)   \; ,
					\\
					\label{eq:css_sr}
			C_{ss} & \simeq 3H^2\left( \eta_{ss}-\frac{1}{3}\eta_{\sigma s}^2\right)  \; .
			\end{align}
		\end{subequations}
	We can proceed similarly with respect to the coefficient $V_{,s}/\dot\sigma$,
	multiplying the first derivatives of the perturbations
	as \mbox{$H\eta_{\sigma s}(1+\varepsilon/3)$}.

          One can now remove the first derivatives from
	the system of equations
	\eqref{eq:eom_usigma} and \eqref{eq:eom_us}
	by performing a change of basis.
	We introduce the rotated field perturbations
	\be
	\left( \begin{array}{c} w_1 \\ w_2 \end{array} \right) = \mathbf{R}^{-1} \left( \begin{array}{c} u_\sigma \\ u_s \end{array} \right)
	\ee
        where the rotation matrix is defined in terms
	of an angle $\vartheta$
	\be
	\mathbf{R} \equiv
	\left( \begin{array}{ccc}
					\cos \vartheta & & -\sin\vartheta  \\
					\sin\vartheta & & \cos \vartheta  \end{array} \right) \ \ ,
		\label{eq:matrixR}
	\ee	
	with
	\be
	\label{eq:dtheta}
	 \dot\vartheta=\frac{V_{,s}}{\dot\sigma} \, .
	\ee
          The equations of motion for the perturbations $w_1$
          and $w_2$ around the time of horizon crossing are,
          to next-order in the slow-roll parameters
          \be
          \left( \begin{array}{c} w_1'' \\ w_2'' \end{array} \right)+\left[ \left(k^2-\frac{a''}{a}\right)\mathbf{1} +\frac{1}{\tau^2}\mathbf{R}^{-1}\mathbf{Q}\mathbf{R} \right]  \left( \begin{array}{c} w_1 \\ w_2 \end{array} \right) = 0 \, ,
          \ee
          where
          \be
          \mathbf{Q} = \left( \begin{array}{cc} 3\eta_{\sigma\sigma}-6\varepsilon+10\varepsilon\eta_{\sigma\sigma} & 3\eta_{\sigma s}+8\varepsilon\eta_{\sigma s}\\  3\eta_{\sigma s}+8\varepsilon\eta_{\sigma s} & 3\eta_{ss}+6\varepsilon\eta_{ss} \end{array} \right) \, .
          \ee
          The assumed pattern of the slow-roll parameters justifies treating them as
          constant in the course of a few e-folds, around the time when the mode of interest with
          a comoving wave number $k_{\star}$ crosses the Hubble radius,
          that is, $k_{\star}=a_{\star}H_{\star}$.
          We can therefore
	replace the slow-roll quantities by their values \textit{at}
	Hubble radius crossing. Moreover, $\vartheta$ can also be replaced by
	its value $\vartheta_\star$ at Hubble radius crossing,
	and we have the freedom of
	choosing the constant term in
	the solution of (\ref{eq:dtheta}) so that the matrix
	$\mathbf{R}^{-1}\mathbf{Q}\mathbf{R} $ is diagonal.
         We can then write the equations of motion for
	the perturbations $w_1$ and $w_2$ as
	\be
	\left( \begin{array}{c}
					w''_{1}  \\
					w''_{2}
		   \end{array} \right) + \bigg[k^2 \mathbf{1} -\dfrac{1}{\tau^2}
		   										\left( \begin{array}{cc}
														2+\lambda_{1\star} & 0  \\
														0 & 2+\lambda_{2\star}
									    	 			  \end{array} \right)
		     \bigg]
		  \left( \begin{array}{c}
					w_{1}  \\
					w_{2}
		   \end{array} \right)  = 0 \ \ ,
		\label{eq:eomw_Ap}
	\ee	
	where we used
	$a''/a=(2-\varepsilon) (aH)^2$.
	Also, when writing Eq.(\ref{eq:eomw}), we used the exact relation fact 
	$V_s/\dot\sigma = H\eta_{\sigma s}/(1-\varepsilon/3)$.
	The linear combinations
	\be
	\lambda_{A\star}=3\varepsilon_{\star}+20\varepsilon_{\star}^2+8\varepsilon_{\star}\eta_{\sigma\sigma\star} - \tilde{\lambda}_{A\star}
	\ee
	where $A\in\{1,2\}$,
	can be expressed in terms of
	the angle $\vartheta_{\star}$ as
		\begin{subequations}
	\begin{align}
		(\tilde{\lambda}_{1\star}-\tilde{\lambda}_{2\star}) \sin(2\vartheta_{\star}) & =
							6\eta_{\sigma s\star} +16\varepsilon_{\star}\eta_{\sigma\sigma\star}\ \ , \\
		(\tilde{\lambda}_{1\star}-\tilde{\lambda}_{2\star}) \cos(2\vartheta_{\star}) & =
		3\eta_{\sigma\sigma\star} -3 \eta_{ss\star} -6\varepsilon_{\star} +10\varepsilon_{\star}\eta_{\sigma\sigma\star} +6 \varepsilon_{\star}\eta_{ss\star}  \; .
			\end{align}
			\label{eq:relations}
		\end{subequations}	
\noindent	 The relations
		\begin{subequations}
	\begin{align}
		\tilde{\lambda}_{1\star}+\tilde{\lambda}_{2\star}  & =
							 3\eta_{\sigma\sigma\star} +3\eta_{ss\star} -6\varepsilon_{\star}  +10\varepsilon_{\star}\eta_{\sigma\sigma\star}
+6 \varepsilon_{\star}\eta_{ss\star} \, , \\
		\tilde{\lambda}_{1\star}^2 + \tilde{\lambda}_{2\star}^2 & \simeq 9(\eta_{ss\star}^2
+2\eta_{\sigma s\star}^2+\eta_{\sigma\sigma\star}^2-4\varepsilon_{\star}\eta_{\sigma\sigma\star}+4\varepsilon_{\star}^2) \, ,
		\label{eq:lambdasquare}
			\end{align}
			\label{eq:other_relations}
		\end{subequations}	
\noindent are also useful for obtaining the two-point correlations of the curvature and isocurvature modes.
	With the assumed hierarchy	
	$\varepsilon,\,|\eta_{\sigma\sigma}|,\,|\eta_{\sigma s}|\ll|\eta_{ss}|<1$
	the results above can be further simplified:
	\begin{subequations}
	\begin{align}
	\tilde{\lambda}_1 & = 3\eta_{\sigma\sigma\star}-6\varepsilon_{\star}+10\varepsilon_{\star}\eta_{\sigma\sigma\star}-9\eta_{\sigma s\star}^2 \, , \; \\
	\tilde{\lambda}_2 & = 3\eta_{ss\star}+6\varepsilon_{\star}\eta_{ss\star} \, . \;
	\end{align}
	\end{subequations}
Note that all the previous equations have been expanded uniformly to
next-order in the slow-roll parameters. However, for the case studied in \S\ref{sec:num_hyb_intro},
only terms quadratic in the largest slow-roll parameter, $\eta_{ss}$, are relevant for maintaining
the required accuracy in the evolution of perturbations.  The only quadratic contributions to the
equations of motion \eqref{eq:eomw} arise from quadratic terms in $\lambda_A$.
The case studied in \S\ref{sec:num_2} has all the slow-roll parameters much smaller than 1
at Hubble radius crossing, therefore the standard results apply.

	\bibliographystyle{JHEPmodplain}
         \bibliography{paper}{}

\end{document}